\documentclass{jpp}
\usepackage{graphicx}
\usepackage{epstopdf, epsfig}

 \usepackage{amssymb}
 \usepackage{bigints}
 \usepackage{amsmath}
\usepackage[utf8x]{inputenc}
 \usepackage{empheq}
 \usepackage{framed}
 \usepackage{color}
 \usepackage{enumerate}
 \usepackage{ulem}
 \usepackage{bm}
 \usepackage{appendix}
 \usepackage{multicol}
 \numberwithin{equation}{section}
 \usepackage{upgreek}
 \usepackage{textgreek}
\usepackage{color}
\usepackage{scalerel}
\usepackage{stackengine,wasysym}
\usepackage{xfrac}    

\newcommand{\be}{\begin{equation}}
\newcommand{\ee}{\end{equation}}
\newcommand{\ba}{\[\begin{aligned}}
\newcommand{\ea}{\end{aligned}\]}
\newcommand{\bea}{\begin{eqnarray}}
\newcommand{\eea}{\end{eqnarray}}
\newcommand{\beann}{\begin{eqnarray*}}
\newcommand{\eeann}{\end{eqnarray*}}
\newcommand{\bs}{\begin{split}}
\newcommand{\es}{\end{split}}

\newcommand*{\cF}{\mathcal{F}}
\newcommand*{\cG}{\mathcal{G}}

\newcommand*{\cJ}{\mathcal{J}}

\newcommand*{\ep}{\epsilon}

\newcommand*{\B}{\bm{B}}

\newcommand*{\dx}{\partial_x}
\newcommand*{\dy}{\partial_y}
\newcommand*{\dz}{\partial_z}

\newcommand*{\dl}{\bm{\nabla}}
\newcommand*{\del}{\partial}
\newcommand*{\BD}{\bm{B}\cdot\bm{\nabla}}
\newcommand*{\BOD}{\BO\cdot\dl}

\newcommand*{\av}[2]{{\left\langle{#1}\right\rangle}_{#2}}
\newcommand*{\aav}[2]{\llangle{#1}\rrangle_{#2}}

\newcommand*{\zh}{\bm{\hat{z}}}

\newcommand*{\lbr}{\left(}
\newcommand*{\rbr}{\right)}

\newcommand*{\psibar}{\overline{\psi}}
\newcommand*{\alphabar}{\overline{\alpha}}

\newcommand*{\aleft}{\tilde{\alpha}}

\newcommand*{\One}[1]{{#1}^{(1)}}

\newcommand*{\alphaO}{\alpha^{(0)}}
\newcommand*{\pO}{p^{(0)}}
\newcommand*{\psiO}{\psi^{(0)}}

\newcommand*{\PhiO}{\Phi^{(0)}}

\newcommand*{\BO}{\B^{(0)}}
\newcommand*{\mBO}{B^{(0)}}
\newcommand*{\FO}{F^{(0)}}

\newcommand*{\alphaOne}{\alpha^{(1)}}

\newcommand*{\psiOne}{\psi^{(1)}}
\newcommand*{\psibarOne}{\psibar^{(1)}}
\newcommand*{\alphabarOne}{\alphabar^{(1)}}
\newcommand*{\alphabarPOne}{\alphabar^{(1)'}}

\newcommand*{\phiOne}{\phi^{(1)}}
\newcommand*{\BOne}{\B^{(1)}}

\newcommand*{\FOne}{F^{(1)}}
\newcommand*{\GOne}{G^{(1)}}
\newcommand*{\HOne}{H^{(1)}}
\newcommand*{\fOne}{f^{(1)}}
\newcommand*{\gOne}{g^{(1)}}
\newcommand*{\PiOne}{\Pi^{(1)}}
\newcommand*{\xOne}{x^{(1)}}

\newcommand*{\alphaTwo}{\alpha^{(2)}}

\newcommand*{\psiTwo}{\psi^{(2)}}
\newcommand*{\psibarTwo}{\psibar^{(2)}}
\newcommand*{\alphabarTwo}{\alphabar^{(2)}}
\newcommand*{\alphabarPTwo}{\alphabar^{(2)'}}
\newcommand*{\PhiTwo}{\Phi^{(2)}}
\newcommand*{\phiTwo}{\phi^{(2)}}

\newcommand*{\HTwo}{H^{(2)}}

\newcommand*{\chiTwo}{\chi^{(2)}}
\newcommand*{\xTwo}{x^{(2)}}

\newcommand*{\xThree}{x^{(3)}}
\newcommand*{\PhiThree}{\Phi^{(3)}}
\newcommand*{\alphaThree}{\alpha^{(3)}}
\newcommand*{\alphabarThree}{\alphabar^{(3)}}

\newcommand*{\xFour}{x^{(4)}}
\newcommand*{\PhiFour}{\Phi^{(4)}}

\newcommand*{\FxOne}{\supsub{F}{x}{1}}
\newcommand*{\FyOne}{\supsub{F}{y}{1}}
\newcommand*{\FzOne}{\supsub{F}{z}{1}}

\newcommand*{\alphaOy}{\alpha^{(0)}_{,y}}
\newcommand*{\alphaOz}{\alpha^{(0)}_{,z}}
\newcommand*{\PhiOy}{\Phi^{(0)}_{,y}}
\newcommand*{\PhiOz}{\Phi^{(0)}_{,z}}

\newcommand*{\xcy}{x_{,y}}
\newcommand*{\xcz}{x_{,z}}
\newcommand*{\xcpsi}{x_{,\Psi}}

\newcommand*{\Phicy}{\Phi_{,y}}
\newcommand*{\Phicz}{\Phi_{,z}}
\newcommand*{\Phicpsi}{\Phi_{,\Psi}}

\newcommand*{\supN}[2]{{#1}^{({#2})}}
\newcommand*{\supsub}[3]{{#1}_{#2}^{({#3})}}
\newcommand*{\supb}[2]{\overline{{#1}}^{({#2})}}
\newcommand*{\supbsub}[3]{\overline{{#1}}_{#2}^{({#3})}}
\newcommand*{\supbPsub}[3]{\overline{{#1}}_{#2}^{({#3})'}}

\newcommand*{\supt}[2]{{\widetilde{{#1}}^{({#2})}}}

\newcommand*\reallywidetilde[1]{\ThisStyle{%
  \setbox0=\hbox{$\SavedStyle#1$}%
  \stackengine{-.1\LMpt}{$\SavedStyle#1$}{%
    \stretchto{\scaleto{\SavedStyle\mkern.2mu\sim}{.5467\wd0}}{.7\ht0}%
  }{O}{c}{F}{T}{S}%
}}

\def\test#1{$%
  \reallywidetilde{#1}\,
  \scriptstyle\reallywidetilde{#1}\,
  \scriptscriptstyle\reallywidetilde{#1}
$\par}
\makeatletter
\newsavebox{\@brx}
\newcommand{\llangle}[1][]{\savebox{\@brx}{\(\m@th{#1\langle}\)}%
  \mathopen{\copy\@brx\mkern2mu\kern-0.9\wd\@brx\usebox{\@brx}}}
\newcommand{\rrangle}[1][]{\savebox{\@brx}{\(\m@th{#1\rangle}\)}%
  \mathclose{\copy\@brx\mkern2mu\kern-0.9\wd\@brx\usebox{\@brx}}}
\makeatother

\newtheorem{theorem}{Theorem}[section]

\shorttitle{Vacuum magnetic fields} 

\title{ Low-shear three-dimensional equilibria and vacuum magnetic fields with  flux surfaces}

\shortauthor{Wrick Sengupta, Harold Weitzner}
\author{Wrick Sengupta, Harold Weitzner}

\affiliation{Courant Institute of Mathematical Sciences, New York University, New York, New York 10012, USA}

\begin{document}

\maketitle

\begin{abstract}
 Stellarators are generically small current and low plasma beta ($\beta= p/B^2\ll1$) devices. Often the construction of vacuum magnetic fields with good magnetic surfaces is the starting point for an equilibrium calculation. Although in cases with some continuous spatial symmetry flux functions can always be found for vacuum magnetic fields, an analogous function does not, in general, exist in three dimensions. This work examines several simple equilibria and vacuum magnetic field problems with the intent of demonstrating the possibilities and limitations in the construction of such states. Starting with a simple vacuum magnetic field with closed field lines in a topological torus (toroidal shell with a flat metric), we obtain a self-consistent formal perturbation series using the amplitude of the nonsymmetric vacuum fields as a small parameter. We show that systems possessing stellarator symmetry allow the construction order by order. We further indicate the significance of stellarator symmetry in the amplitude expansion of the full ideal magnetohydrodynamics (MHD) problem as well. We then investigate the conditions that guarantee neighboring flux surfaces given the data on one surface, by expanding in the distance from that surface. We show that it is much more difficult to find low shear vacuum fields with surfaces than force-free fields or ideal MHD equilibrium. Finally, we demonstrate the existence of a class of vacuum magnetic fields, analogous to ``snakes'' observed in tokamaks, which can be expanded to all orders. 
 
\end{abstract}

\section{Introduction \label{sec:intro}}

Designing vacuum fields with nested flux surfaces is highly relevant to modern stellarators. Understanding the properties of such fields could provide insights for better optimization of stellarator equilibrium. However, several mathematical challenges in finding such fields arise due to the mixed nature of the partial differential equations (PDEs) for vacuum fields with surfaces. A vacuum magnetic field satisfies Laplace's equation. If one further requires that the vacuum magnetic field in a generalized torus also have flux surfaces, one can reformulate the problem as a third order system of PDEs (c.f. discussion in section \ref{sec:Cauchy_vac} Eq. \ref{sysa10a11a12}). Although one may readily create such a formalism, there is no assurance that non-trivial nonsymmetric solutions exist. As \cite{grad1967toroidal} indicated, there is no complete mathematical theory for this problem. Except for highly symmetric cases, a global separation between the elliptic (imaginary) and hyperbolic (real) characteristics is not possible \citep{Garabedian2012MHDstellarator}. In a non-symmetric three-dimensional system, lack of compatibility between the elliptic and the hyperbolic part leads to resonances near rational surfaces where the real characteristic generally generate singularities \citep{Garabedian2012computational}. Near such regions, flux surfaces typically do not exist, and magnetic islands and stochastic regions are often present.

The difficulties of obtaining vacuum fields with nested flux surfaces can also be understood from a dynamical systems point of view since magnetic field line flows are Hamiltonian systems with $1\sfrac{1}{2}$ degrees of freedom \citep{caryLittlejohn1983Hamiltonian_B}. The axisymmetric system is integrable, but three-dimensional field-line flow is in general non-integrable. For Hamiltonians close to integrable ones, KAM theory \citep{lichtenberg1992regular} is applicable, provided the magnetic shear is finite and non-zero everywhere. The finite non-zero magnetic shear condition is known as the ``Kolmogorov non-degeneracy'' condition \citep{chierchia2010kolmogorov,hanssmann2011non}. For any generic perturbation of the non-degenerate system, magnetic islands and stochastic field lines are obtained near most surfaces with a rational rotation transform. The highly irrational surfaces that persist for large enough perturbations, typically show self-similar fractal behavior \citep{hudsonKrauss20173D_cont_B,morrison2000magnetic}. MHD equilibrium with such fractal solutions has been recently studied in \cite{krausHudson2017fractal_pressure}. Hamiltonians that differ non-perturbatively from integrable ones are of course much more difficult to understand. 

In studies of plasma systems, we often encounter situations where the magnetic shear is weak, such as the Wendelstein stellarators, or the shear has zeros as in the case of reverse field pinch. Mathematically the field line flows of such ``weakly non-degenerate" systems can be modeled by the non-twist maps \citep{delcastillo1996nontwist_chaos,morrison2000magnetic}. The finite dimensional classic KAM theorem does not directly apply to such maps \citep{abdullaev2006construction} and generalizations of KAM \citep{delsham_delaLlaves2000kAM_Greene, gonzalez_delaLlave2014singularity} must be considered. When a small non-integrable perturbation is added to an integrable non-twist map, new characteristics such as persistence of certain shearless-orbits are found. Other features like self-similarity of the persistent torus with highly irrational rotation transform, and formation of magnetic islands and stochastic behavior of field lines near rational surfaces, are shared with the standard KAM theory \citep{delsham_delaLlaves2000kAM_Greene}. While a lot of the current work on non-twist maps has been focused on the most irrational or the noble-winding numbers \citep{delcastillo1996nontwist_chaos}, the behavior near low-order rational surfaces has also been explored \citep{morozov2002degenerate, karabanov2014degenerate}. 

It is known that the behavior of a low-shear magnetic field system can be markedly different in the neighborhood of closed field lines (i.e., rational surfaces when they persist) or generic ergodic surfaces \citep{firpo2011study}. Numerous experimental results from Wendelstein VII-A/AS \citep{jaenicke1993w7ASvacuum,hirsch2008majorW7AS,brakel2002energytransp_rational_iota_W7AS,brakel1997W7AS_EB_shear} and numerical results \citep{wobig1987localized_pert_w7A} clearly support the idea that optimum confinement is usually found close to the narrow resonance-free zones which exist in the direct vicinity of certain low-order rational surfaces. For arbitrary perturbations, the islands on these surfaces are
found \citep{wobig1987localized_pert_w7A} to be exponentially small in size. A dynamical systems explanation for such a behavior is seen in
V.I.Arnold's work \citep{arnol1963smalldiv,arnold_dynamical_iii} who referred to the low-shear systems as ``properly degenerate systems". We have included a discussion of KAM theory for such systems in \ref{App:AppendixA}. In short, Arnold showed that low-shear systems are in some sense ``more integrable" than the usual perturbed system and as long as magnetic shear is small but non-zero invariant tori continue to exist for systems of $1\sfrac{1}{2}$ degrees of freedom. The persistence of tori in lower dimensional low-shear Hamiltonian systems is closely related to the concept of ``perpetual stability'' \citep{chierchia2010kolmogorov}.
 
The application of low-shear KAM theory is suggestive, but not necessarily definitive, unless the theory can assert that for any variation of the magnetic field structure surfaces are destroyed. Qualitatively the theory suggests that the region of a surface breakup, if any, is smaller than for the case of a background with shear. Thus, even if the formal expansions were not to converge, they would possibly have a richer surface structure. 

In stellarator literature, several numerical techniques have been developed to reduce the stochastic regions. These techniques reduce magnetic resonances by adding extra harmonic fields and support the idea that vacuum fields with good nested surfaces may exist in isolated cases \citep{cary1982vacuum, Caryhanson1984elimination,dommaschk1986representations,merkel1987solution,hudson2001reduction}.
Formal analytic expansions of equilibria in a torus have also been obtained with different expansion schemes. For example, large aspect ratio  expansions  of stellarator equilibrium have been studied extensively \citep{greene1961determination,freidberg_idealMHD,strauss1980stellarator,dobrott1971magnetic}. Although these expansions provide physical intuition, it is not clear if the resonances can be in general eliminated to all orders. On the other hand, formal expansions in the amplitude of three-dimensional fields about a closed magnetic field line system have been carried out to all orders by \cite{spiesLortz1971asymptotic}. This latter work, however, does not insert either equilibrium or vacuum magnetic field and requires only that the magnetic field be divergence-free. Several authors \citep{grad1973magnetofluid,strauss1981limiting_beta_vacuum} have highlighted the stability of low-shear systems.

Poincare's theorem on separatrix spitting \citep{holmes1988exponentially} makes it clear that if we seek nested surfaces everywhere by perturbing an integrable system, the perturbation cannot be generic. Unlike the axisymmetric case where the profile can be arbitrarily chosen, non-symmetric smooth rotation transform profile needs to be self-consistent with the magnetic fields to avoid small-divisor singularities on the rational surfaces \citep{newcomb1959magnetic,grad1967toroidal,hudsonKrauss20173D_cont_B}. 
As discussed earlier, results from dynamical systems, experiments and numerics suggest that low-shear systems in the vicinity of low-order rational surfaces could support a highly constrained but smooth vacuum or MHD equilibria with nested surfaces.

In this work, we construct low-shear nested vacuum magnetic fields, assuming closed magnetic fields to lowest order. Our goal is not to study the response to a generic perturbation but to look for particular classes of perturbation for which resonance conditions can always be satisfied. Previous work by Weitzner \citep{weitzner2014ideal,weitzner2016expansions} suggest that in low-shear systems, ideal MHD equilibrium formal expansions can be carried out to all orders around low order rational surfaces with closed field lines for both perturbative and non-perturbative deviations from axisymmetry. The low-shear closed field line systems provide unique flexibility in avoiding the singular divisor problem as will be shown in details later. The current work extends the methodology to vacuum fields and compares and contrast the expansions for ideal MHD equilibrium and vacuum fields with surfaces. 

The paper is structured as follows. In section \ref{sec:2D}, we discuss the basic mathematical structure of three-dimensional vacuum fields that possess nested flux surfaces. In section \ref{sec:straightB0}, we perform an asymptotic amplitude expansion similar to \citep{weitzner2014ideal,spiesLortz1971asymptotic}. We find that in systems with stellarator symmetry the expansion can be carried to all orders. Next, in section \ref{sec:AmpMHD}, we  revisit the amplitude expansion in full ideal MHD equilibrium \citep{weitzner2014ideal}. We find a similarly important role played by stellarator symmetry.   
In section \ref{sec:Cauchy_vac}, we tackle the problem of the existence of nested non-symmetric flux surfaces for vacuum fields with non-perturbative amplitude. Assuming data on a flux surface, we construct solutions in the neighborhood of the surface following \cite{weitzner2016expansions}. We show the existence of highly constrained solutions which are analogous to a tokamak ``snake-like" equilibrium \citep{gill1992snake,cooper2011jetsnake,senguptaHasssam2017subalf}. 
We discuss our findings and conclude in section \ref{sec:conclusions}.

\section{Vacuum magnetic fields with nested flux surfaces}
\label{sec:2D}

We now discuss the mathematical structure of magnetic fields with nested flux surfaces. A divergence-free magnetic field may always be expressed in terms of two Clebsch variables $(\psi,\alpha)$ as 
 \begin{align}
 \B = \dl\psi\times \dl\alpha
 \label{ClebschB}
 \end{align}
 They both satisfy a homogeneous magnetic differential equation since
\begin{subequations}
\begin{align}
 \BD\psi=& 0 \label{Clebschpsi}\\
 \BD\alpha=& 0 \label{Clebschalpha}
 \end{align}
 \label{ClebschMDE}
 \end{subequations}
The Clebsch variables are multivalued in general in a torus. If we assume that $\B$ is to have nested magnetic surfaces  then we may choose $\psi$ to be a single-valued flux function while $\alpha$ need not be single valued in a torus. However, $(\psi, \alpha)$ must be chosen so that $\B$ is single-valued. Further, since $\B$ is curl-free we may write 
\begin{align}
\B=\dl\Phi,
\label{vacuumB}
\end{align} 
where $\Phi$ is in general multivalued. Finally, the three unknown functions $(\Phi,\psi,\alpha)$ must satisfy the vector equation
\begin{align}
\dl\Phi = \dl\psi\times \dl\alpha
\label{Bveceqn}
\end{align}
It is clear that $\Phi$ must satisfy a three-dimensional Laplace's equation
\begin{align}
\Delta \Phi=0.
\label{Laplacian_Phi}
\end{align}
As is well known \citep{grad1967toroidal, weitzner2016expansions}, in a solid toroidal domain labeled by flux surfaces $\psi$ and  poloidal and toroidal angles $(\theta,\varphi)$ respectively, the multivalued potentials take on the following form
\begin{align}
\Phi(\psi,\theta,\varphi)=& F \theta + G \varphi +\widetilde{\Phi}(\psi,\theta,\varphi) \nonumber\\
\alpha(\psi,\theta,\varphi)=& f(\psi) \theta + g(\psi) \varphi +\widetilde{\alpha}(\psi,\theta,\varphi)
\label{formPhiAlf}
\end{align}
where the functions $(\widetilde{\Phi},\widetilde{\alpha})$ are doubly periodic functions of the two angles. $(F,G)$ are related to the poloidal and the toroidal flux respectively as can be seen by computing the flux $\oint \B\cdot \bm{dA}$ through the two sections.  $F$ and $G$ must be constants in order that $\B=\dl \Phi$ is single valued. The flux functions $(f,g)$ are related to the rotation transform by 
 \begin{align}
 \iota(\psi)=\frac{g(\psi)}{f(\psi)}
 \label{iota}
 \end{align} Although we can satisfy the conditions (\ref{Clebschpsi}) locally, finding a global solution of the form given by Eqn. (\ref{formPhiAlf}) is highly non-trivial.

\section{Amplitude expansion near a torus with a closed field line}
\label{sec:straightB0}
As mentioned earlier, the general question of the existence of vacuum magnetic fields with non-symmetric three-dimensional magnetic flux surface is still an open problem. However, one could take a perturbative approach and look for solutions near one system which already possess nested flux surfaces. As discussed in section IX of \citep{taylor_hastie}, analytical progress can be made if we regard the field as composed of a large field with closed field lines everywhere, and a smaller additional perturbative field which produces an irrational rotation transform and the ergodic behavior on a flux surface. This method of decomposition of magnetic fields is especially useful in describing a low shear system near a rational surface. We shall now present the details of a perturbation analysis where the expansion parameter is the amplitude of the added perturbations to the closed field lines. The goal is to solve (\ref{Clebschpsi}) for $\psi$ order by order. In each order $n$ we obtain an inhomogeneous magnetic differential equation for $\psi^{(n)}$ driven by lower order quantities. The solvability condition \citep{newcomb1959magnetic} for this equation then constrains the lower order.

We define a toroidal shell (referred to as a topological torus in \cite{weitzner2016expansions}) where $y$ and $z$ are the two angles extending from 0 to $2\pi$ and the thickness in $x$ is assumed to be finite. We assume Euclidean metric ($ds^2= dx^2 +dy^2 +dz^2$) on the shell. The lowest order magnetic field is assumed to be a uniform field of unit strength in the $z$ direction (given by the unit vector $\zh$). The lowest order flux surface $\psi_0$ is chosen to be the $x=0$ plane, while the lowest order field line label $\alpha_0$ is chosen to be the $y=0$ plane.  Therefore, 
 \be \B_0=\zh,\quad\psi_0=x,\quad\alpha_0=y. \label{lowest0}\ee
This satisfies (\ref{ClebschMDE}) conditions trivially. The potential $\Phi$ and the field line label $\alpha$ are multivalued in a toroidal domain. However, the gradients of these functions need to be single valued since physical quantities such as the magnetic field need to satisfy periodic boundary conditions in $y$ and $z$ directions. Taking into account the general structure as given by (\ref{formPhiAlf}), we now expand as follows.
    \begin{align}
        \phi&= \epsilon\: \phi^{(1)}+ \epsilon^2\: \phi^{(2)}+..,\quad \psi=x + \epsilon \psi^{(1)}+\epsilon^2 \psi^{(2)}+..,\quad \aleft=\ep \aleft^{(1)}+..\nonumber\\
 F&=\epsilon F^{(1)}+\epsilon^2 F^{(2)}+..,\quad G=1 + \epsilon G^{(1)}+..,\:\: g=\epsilon g^{(1)}+..\quad f=1+\ep f^{(1)} 
 \label{Ampexpansions}
  \end{align}
  $\phi,\psi,\aleft$ are single valued and doubly periodic in $(y,z)$. From (\ref{Laplacian_Phi}) we get at each order 
 \begin{align}
 \Delta \phi^{(n)}=0
 \label{delphiN}
 \end{align}
Let us define the following operations $$\av{A}{u} \equiv \oint \frac{du}{2 \pi} A, \quad \aav{A}{u}\equiv A-\av{A}{u}$$
where $u$ can be the angle $y$ or $z$.
To first order in $\epsilon$ the condition (\ref{Clebschpsi}) implies
\begin{align}
\del_z \psi^{(1)}+\phi^{(1)}_{,x}=0.
\label{ep1storder}
\end{align}
Since $\psiOne$ must be a single valued function of $y$ and $z$, averaging over the angle $z$ we get the following consistency condition
\begin{align}
\dx \av{\phi^{(1)}}{z} =0
\label{1storderNewcomb}
\end{align}
As a consequence, we can express $\phiOne$ in terms of a harmonic function $\HOne$ such that
\begin{align}
\phiOne=\dz \HOne, \quad \Delta \HOne=0 \quad \text{and} \quad \av{\HOne}{z}=0
\label{phi1H1}
\end{align}
The last condition follows from the fact that any $z$-independent component of $\HOne$ does not contribute to $\phiOne$ and we are free to set it to zero. We can solve (\ref{ep1storder}) for $\psiOne$
\begin{align}
\psi^{(1)}={\psibar}^{(1)} -H^{(1)}_{,x} \quad \text{where}\quad  \psibar^{(1)}=\psibar^{(1)}(x,y)=\av{\psiOne}{z}
\label{eppsi1sol}
\end{align}
At this stage ${\psibar}^{(1)} $ is a ``free-function'' and will be fixed later by demanding solvability of higher order equations. When Eqn. (\ref{Clebschpsi}) is taken to second order we obtain after some straight forward algebra
\begin{align}
 \phiTwo_{,x}+&    \del_z\lbr \psiTwo+(\GOne+\HOne_{,x}\del_x +\HOne_{,y}\del_y)\psiOne\rbr \nonumber\\
     &=\del_x\lbr\HOne_{,x}\psiOne_{,z}\rbr+\del_y\lbr\HOne_{,y}\psiOne_{,z}-\FOne\psiOne\rbr 
     \label{ep2ndorder}
\end{align}
We could try a solution for $\phiTwo$ of the form $\phiTwo=\del_z \HTwo+\overline{\phiTwo}(x,y)$, with $\HTwo$ being a harmonic function. Upon averaging Eqn. (\ref{ep2ndorder}) with respect to $z$ we then get an equation for $\overline{\phiTwo}$ in terms of $(\HOne,\psiOne)$. However, $\phiTwo$ need to also satisfy Laplace's equation  (\ref{delphiN}). This implies that the $z-$independent part of $\phiTwo$ can not be arbitrarily fixed. Thus the only consistent solution is  \be \phiTwo=\del_z \HTwo \label{epphi2sol}\ee
Since all the quantities within the $z$ derivative in Eqn. (\ref{ep2ndorder}) are single valued functions, averaging (\ref{ep2ndorder}) with respect to $z$ annihilates those terms. We also find that the $\dx$ term in (\ref{ep2ndorder}) vanishes due to Eqn. (\ref{eppsi1sol}) resulting in an expression for $\psibar^{(1)}$
 \be \FOne \psibar^{(1)}= -\oint \frac{dz}{2\pi}\HOne_{,y}\HOne_{,xz} \label{psi1bar1eqn}
 \ee
 Provided $\psibar^{(1)}$ satisfies (\ref{psi1bar1eqn}) we see that the periodicity in $z$ constraint is satisfied. Next we proceed to solve for $\psiTwo$. Defining $$\chiTwo\equiv \dy\int dz\: \lbr\FOne \HOne_{,x}+\aav{\HOne_{,y}\psiOne_{,z}}{z}\rbr,$$ 
 we can now solve (\ref{ep2ndorder}) for $\psiTwo$
 \be \psiTwo=\psibarTwo +\chiTwo -\HTwo_{,x}+\GOne \HOne_{,x}-\aav{(\HOne_{,x}\del_x +\HOne_{,y}\del_y)\psiOne}{z} \label{eppsi2sol}\ee

We can proceed to the general $n^{th}$ order. Once again no consistent solution is found unless 
\be
\supN{\phi}{n}=\del_z \supN{H}{n}
\label{eq:PhiN}
\ee
The surface condition (\ref{Clebschpsi}) to this order provides an equation for $\supN{\psi}{n}$ 
   \begin{align}
     \del_z\lbr \supN{\psi}{n}+\supsub{H}{,x}{n}+\sum_{i}^{n-1}(\supN{G}{i}+{\supsub{H}{,x}{i}}\del_x +{\supsub{H}{,y}{i}}\del_y){\supN{\psi}{n-i}}\rbr \nonumber\\
     =\del_x\lbr\sum_{i}^{n-1}{\supsub{H}{,x}{i}}\supsub{\psi}{,z}{n-i}\rbr +\del_y\sum_{i}^{n-1}\lbr{\supsub{H}{,y}{i}}\supsub{\psi}{,z}{n-i}-{\supN{F}{i}}{\supN{\psi}{n-i}}\rbr 
     \label{epNdorder}
    \end{align}
Averaging over $z$ and demanding periodicity in $z$ as before, we get   
\be 
\del_y\sum_{i}^{n-1}\lbr\av{{-\supsub{H}{,y}{i}}\supsub{\psi}{,z}{n-i}}{z}+{\supN{F}{i}}{\supb{\psi}{n-i}}\rbr=\del_x\av{\sum_{i}^{n-1}{\supsub{H}{,x}{i}}{\supsub{\psi}{,z}{n-i}}}{z}
\label{ZavgepNorder}
\ee 
Since we also need periodicity in $y$, we average (\ref{ZavgepNorder}) over $y$ and obtain the following constraint 
\begin{align}
\sum_{i}^{n-1}\oint \frac{dy}{2\pi}\oint \frac{dz}{2\pi}{\supsub{H}{,x}{i}}{\supsub{\psi}{,z}{n-i}} =\text{constant}
\label{mainconstraint}
\end{align}
It can be checked that the integral in (\ref{mainconstraint}) identically vanishes at least to $O(\ep^4)$. For higher $n$, we can give no guarantee that this will be satisfied. More work is needed to see if this is automatically satisfied due to the Hamiltonian structure of the magnetic field lines. It will be shown later in section \ref{subsec:stellsym}, that for systems with stellarator symmetry, this condition can be satisfied.  As a consequence, we can define a function $\supb{Q}{n}(x,y)$ such that
\begin{align}
\sum_{i=1}^{n-1}\oint \frac{dz}{2\pi}\supsub{H}{,x}{i}\supsub{\psi}{,z}{n-i} =\dy \supb{Q}{n}(x,y)+\supsub{Q}{0}{n}
\label{Qbdef}
\end{align}
where $\supb{Q}{n}(x,y)$ is periodic in $y$ and $\supsub{Q}{0}{n}$ is a constant. In systems with stellarator symmetry, we can show that such a $\supb{Q}{n}(x,y)$ indeed exists. In such a system we can solve (\ref{ZavgepNorder}) for $\supb{\psi}{n-1}$ as follows
 \be
 \FOne \psibar^{(n-1)}=-\lbr\sum_{i=2}^{n-1}\supN{F}{i}\supb{\psi}{n-i}\rbr+\del_x \supb{Q}{n}(x,y)+\sum_{i=1}^{n-1}\av{{\supsub{H}{,y}{i}}\supsub{\psi}{,z}{n-i}}{z}\:.
 \label{eppsibNm1}   
 \ee   
 Next we define  
 \be 
 \supN{\chi}{n}\equiv \int dz\:\lbr \del_x\sum_{i=1}^{n-1}\aav{\supsub{H}{,x}{i}\supsub{\psi}{,z}{n-i}}{z} +\del_y\sum_{i=1}^{n-1}\aav{\supsub{H}{,y}{i}\supsub{\psi}{,z}{n-i}-{\supN{F}{i}}{\supN{\psi}{n-i}}}{z} \rbr  \:.
 \label{epchiN}
 \ee
Finally the general solution for $\supN{\psi}{n}$ can be written as
 \begin{align}
    \supN{\psi}{n}(x,y,z)=& \:\supb{\psi}{n}(x,y)-\supsub{H}{,x}{n}+\supN{\chi}{n}-\sum_{i}^{n-1}\supN{G}{i}\aav{\supN{\psi}{n-i}}{z} \label{psingeneralsol}\\
    &\:-  \sum_{i}^{n-1}\aav{{(\supsub{H}{,x}{i}\del_x +\supsub{H}{,y}{i}\del_y){\supN{\psi}{n-i}}}}{z}\nonumber
    \end{align}
By construction periodicity in $y$ and $z$ are maintained to all orders. 
\subsection{Calculation of the field line label $\alpha$}
Similar to solving the magnetic differential equation for $\psi$ we can solve (\ref{Clebschalpha}) $\alpha$ order by order. To first order
\begin{align}
\BOD\One{\alpha}+\B^{(1)}\cdot \dl \alphaO=0
\label{MDEalf1}
\end{align}
For the straight magnetic field we considered earlier, with $$\alpha_0=y,\quad \alphaOne= \fOne y + \gOne z +\supt{\alpha}{1}$$
 We find that (\ref{MDEalf1}) simplifies to $$\del_z\lbr \supt{\alpha}{1} + \del_y \supN{H}{1}\rbr +\lbr \gOne+\FOne\rbr=0$$ This implies that 
 \be \gOne=-\FOne,\:\:\: \supt{\alpha}{1}= -\del_y \HOne 
 \label{alpha1sol}
 \ee
From (\ref{iota},\ref{Ampexpansions}) and (\ref{alpha1sol}), the rotation transform is seen to be $-\FOne$. The functions $(\psibarOne,\alphabarOne)$ are not independent because they are related by 
 \begin{align}
 \B^{(1)}=\dl \psi^{(0)} \times \dl \alpha^{(1)}+\dl \psi^{(1)} \times \dl \alpha^{(0)} =\dl \Phi^{(1)}
 \label{ClebschB1}
 \end{align}
 It can be easily checked that Eqn. (\ref{ClebschB1}) satisfies both the MDEs (\ref{Clebschpsi},\ref{Clebschalpha}). Also note that there is a gauge degree of freedom in Eqn. (\ref{ClebschB1}), we can add any function of $\psiO$ to $\alphaOne$ and any function of $\alphaO$ to $\psiOne$ and keep $\BOne$ unchanged i.e.
 \begin{align}
 \psiOne \rightarrow \psiOne +\cF(\alphaO), \quad  \alphaOne \rightarrow \alphaOne +\cG(\psiO),\quad  \BOne \rightarrow \BOne
 \label{gauge}
 \end{align}
 Finally, we see that 
 \begin{align}
 \psibarOne_{,\psiO} + \alphabarOne_{,\alphaO}=\frac{\BO\cdot\BOne}{{(\mBO)}^2}
 \label{relnInbars}
 \end{align}
 We can proceed to higher orders but once the consistency conditions are satisfied for $\psi$ we can obtain $\alpha$ up to the gauge discussed above. The only issue left to deal with, is the existence of $\supb{Q}{n}(x,y)$. In the following we shall discuss a class of solutions where the constraint (\ref{mainconstraint}) can be satisfied. 

\subsection{A class of perturbative expansion and error analysis}
\label{subsec:stellsym}
We note that if $\supN{H}{i}$ and $\supN{\psi}{j}$ have the same parity under $z$ and $y$ for all $(i,j)$ then the condition is satisfied.
If $\supN{H}{i}(x,y,z)=\supN{H}{i}(x,-y,-z)$ (i.e even under inversion of angles), and so is $ \supN{\psi}{j}(x,y,z)$ for all $(i,j)$, then the integrand in (\ref{mainconstraint}) is odd under exchange of $(y,z)$ integrations and the integral vanishes.
From the general solution $\supN{\psi}{n}(x,y,z)$ it can be checked that this discrete symmetry (stellarator symmetry) is preserved order by order.

Stellarator symmetry \cite{dewar1998stellarator} is formally defined as follows: If S denotes the symmetry operator then its action on scalar $f$ and vector $\bm{F}$ is
 \begin{align*}
S f(x,y,z)=f(x,-y,-z)\\
S \{F_x,F_y,F_z\}=\{-F_x,F_y,F_z\}
\end{align*}
In our context we have shown that if we add a stellarator symmetric three-dimensional harmonic function of the form $\dz H^{(n)}$, it is possible to obtain a formal flux surface to all orders. Stellarator symmetry is only a sufficient but not necessary condition here. There might be other classes of solutions that satisfy (\ref{mainconstraint}). 

We have not carried out a convergence test. Although we are not considering generic perturbations, we can still use the low-shear KAM arguments (given in Appendix \ref{App:AppendixA}) to estimate the region of validity of the expansion. Theorem \ref{thm2} indicates that for a generic perturbation, the measure of surfaces that break up to form islands or regions of stochasticity is exponentially small, i.e., $O(\exp{(-1/\epsilon)})$, which is smaller than any polynomial power of $\epsilon$.  Therefore, even if we fail to satisfy the resonance conditions, the effect of resonance is going to be limited to an exponentially small region near the resonant surface, outside of which we expect the expansion to be valid. Provided we truncate our series at $O(n)$, we shall be able to compute the flux surface up to that order with an error of $O(\epsilon^n)$, but it is not clear how large or small $n$ must be in case the series is asymptotic and divergent. For an optimal $n$, the partial sum approximation can be made exponentially accurate.

 We shall now revisit the amplitude expansion of ideal MHD equilibrium as presented in \citep{weitzner2014ideal}. We shall find that stellarator symmetry is also crucial in obtaining a self-consistent perturbative solution to the full ideal MHD system. 
 
\section{Amplitude expansion in ideal MHD}
\label{sec:AmpMHD}
In this section we return to the construction of non-symmetric ideal MHD equilibrium in a topological torus treated in \citep{weitzner2014ideal}.In that paper shear was assumed to be nonzero. We extend the methodology to address the possibility of zero average magnetic shear to first order in the amplitude of the nonsymmetric magnetic field. We show that a different formulation from that in \citep{weitzner2014ideal} is needed for zero average shear and use a more explicit representation than previously employed. We write the equations of an ideal MHD equilibrium as
\begin{subequations}
\begin{gather}
\dl \cdot \B=0 \label{eq1a}\\
\dl \Pi= \B \cdot \dl \B, \quad \Pi = p+\frac{1}{2}B^2.
\label{eq1b}
\end{gather}
\label{sys1}
\end{subequations}

We observe that in a torus $\Pi$ must be single valued. We take the toroidal shell to be a domain in three dimension and require the equilibrium to be periodic of period $2\pi$ in $y$ and $z$. As in \citep{weitzner2014ideal}, and unlike \citep{weitzner2016expansions} we expand the solution in a formal series in the amplitudes of the components of the magnetic fields. We choose the lowest order field to be a function of $x$ alone, and without loss of generality we take
\begin{align}
\supN{\B}{0}= (0,0,\supN{F}{0}(x)), \quad \text{for which} \quad \supN{\Pi}{0}=\supN{p}{0}+\frac{1}{2}\supN{F}{0}(x)^2= \text{constant}.
\label{eq2}
\end{align}
We may also consider the first order variation of the system \ref{sys1} and we set 
\begin{align}
\supN{\B}{1}=(\supsub{F}{x}{1}(x)\cos{\omega},\supsub{F}{y}{1}(x)\sin{\omega},\supsub{F}{z}{1}(x)\sin{\omega}),\quad \supN{\Pi}{1}=\supN{G}{1}(x)\sin{\omega}
\label{eq3}
\end{align}
where, \be \omega= m y + n z \label{eq4} \ee
with integer $m$ and $n$. Since all the components of $\BOne$ are periodic, the average shear to this order is zero. We obtain readily
\begin{subequations}
\begin{align}
&\del_x\supsub{F}{x}{1}+ m \supsub{F}{y}{1}+n \supsub{F}{z}{1}=0 \label{5a}\\
&\del_x \supN{G}{1} = -n \supN{F}{0} \supN{F}{1}_{x}\label{5b}\\
&m\supN{G}{1} = n \supN{F}{0} \supsub{F}{y}{1} \label{5c}\\
&n\supN{G}{1} = n \supN{F}{0} \supsub{F}{z}{1}+\del_x\supN{F}{0} \:\supsub{F}{x}{1} 
\label{5d}
\end{align}
\label{sys5}
\end{subequations}
So that 
\begin{align}
\lbr m^2 +n^2\rbr \GOne= n (-\FO  \del_x\FxOne + \del_x\FO \FxOne)
\label{eq6}
\end{align}
Thus, the above equations determine $\FxOne$ as a solution of
\begin{align}
n\lbr \frac{1}{{(\FO)}^2}\dx\lbr  {(\FO)}^2 \dx \lbr \frac{\FxOne}{{\FO}}\rbr \rbr-(m^2+n^2) \lbr \frac{\FxOne}{{\FO}}\rbr \rbr=0\:.
\label{eq7}
\end{align}
Provided $n\neq 0$ $\FyOne,\FzOne$ are then given by (\ref{5c},\ref{5d}). When $n=0$ we see from (\ref{5c}) that $\PiOne=0$ so that from (\ref{5d}) $\FxOne=0$. If $m\neq 0$ then both $\FxOne,\FyOne$ are zero but $\FzOne\neq 0$ and therefore
\begin{align}
\B^{(1)}=(0,0,\FzOne)\cos{(m y +\lambda)} \quad \text{for} \quad (m\neq 0,n=0)
\label{eq8}
\end{align}
with $\lambda$ equal to a constant. If $m=n=0$ then 
\begin{align}
\B^{(1)}=(0,\FyOne(x),\FzOne(x)) \quad \text{for} \quad (m=n=0)
\label{eq9}
\end{align}
where the functions in (\ref{eq8},\ref{eq9}) are arbitrary functions of $x$, while $\PiOne$ is a constant in both these cases.

 We can now develop the power series solution to the system. We select fixed values of $m$ and $n$, neither of which is zero for the first order magnetic field. We may add other fields with given $m$ and $n$ in higher orders. In higher order, the system of equation involving $\supN{\B}{N}$, after Fourier analysis will look exactly like the system (\ref{5a},\ref{5b},\ref{5c},\ref{5d}) except that there will be inhomogeneous terms added to them coming from products of lower order terms of index at least one. The system can be solved for $\supN{\B}{N}$ provided only that the corresponding value of $n$ is not zero. When the product of lower order inhomogeneous terms, after Fourier analysis has an index $n$, the system has no solution. Thus, the resonance destroys the equilibrium. Exactly as in many problems in classical mechanics, one can look to eliminate the resonance by modifying the solution in some order by the addition of special lower order ``resonant terms" which suppress the potential singularity. We have seen here such resonant terms exist and are given by (\ref{eq8},\ref{eq9}). We can use such terms to suppress resonant singularities in higher orders. In \cite{weitzner2014ideal} it was argued that if the combination of lower order terms which generate a resonance appear in order $N$, then the addition of resonant field (\ref{eq8},\ref{eq9}) in order $(N-1)$ could cancel the effects of the resonating combinations. Another paper \cite{weitzner2016expansions} carried out a similar calculation for a related problem and validated the basic idea. The basic idea is indeed correct but requires shear to be nonzero. For systems with zero average shear, we show here that we need to include the resonant mode at $O(N-2)$ instead of $O(N-1)$. In the following, we treat separately the cases $(m\neq 0, n=0)$ and $(m=n=0)$. 
 
 We start with the more complex situation $(m\neq 0, n=0) $ present in order $N$ and we add in order $(N-2)$ a resonant magnetic field
 \begin{align}
 \supN{\B}{N-2}=(0,0,\supsub{F}{z}{N-2})\cos{(\mu y +\lambda)},
 \label{10a}
 \end{align}
 where $\mu\neq 0$ and $\lambda$ is an arbitrary constant. We now examine the magnetic field in order $(N-1)$. We must find the new terms in this order which come from the interaction of the order $(N-2)$ resonant fields and order 1 fields with given period 1. We get

 \begin{subequations}
 \begin{align}
 0=&\supN{B}{N-1}_{x,x}+ \supN{B}{N-1}_{y,y}+ \supN{B}{N-1}_{z,z} \label{eq11a}\\
 \supN{\Pi}{N-1}_{,x}=&\FO \supN{B}{N-1}_{x,z}-\frac{n}{2}  \supN{F}{N-2}_{z}\supN{F}{1}_{x}\left( \sin{(\omega-(\mu +\lambda))} + \sin {(\omega+(\mu +\lambda))} \right) \label{eq11b}\\\
  \supN{\Pi}{N-1}_{,y}=&\FO \supN{B}{N-1}_{y,z}+\frac{n}{2}  \supN{F}{N-2}_{z}\supN{F}{1}_{y}\left( \cos{(\omega-(\mu +\lambda))} + \cos {(\omega+(\mu +\lambda))} \right) \label{eq11c}\\\
   \supN{\Pi}{N-1}_{,z}=&\FO \supN{B}{N-1}_{z,z} +\dx\FO \supsub{B}{x}{N-1}\label{eq11d}\\\
   &+\frac{1}{2}\cos{(\omega+(\mu +\lambda))}\lbr \supN{F}{1}_x\dx\supsub{F}{z}{N-2}+\mu \supN{F}{1}_y\supsub{F}{z}{N-2}+n\supN{F}{1}_z\supsub{F}{z}{N-2}\rbr \nonumber \\
  &+\frac{1}{2}\cos{(\omega-(\mu +\lambda))}\lbr \supN{F}{1}_x\dx\supsub{F}{z}{N-2}-\mu \supN{F}{1}_y\supsub{F}{z}{N-2}+n\supN{F}{1}_z\supsub{F}{z}{N-2}\rbr\nonumber
 \end{align}
 \label{sys11}
 \end{subequations}
We see that there are two components to this solution. 
 
 \begin{subequations}
 \begin{align}
 \{{\supN{\Pi}{N-1}}^{\pm},{\supsub{B}{y}{N-1}}^{\pm},{\supsub{B}{z}{N-1}}^{\pm}\}&=\{{\supN{G}{N-1}}^{\pm},{\supsub{F}{y}{N-1}}^{\pm},{\supsub{F}{z}{N-1}}^{\pm}\}\sin{(\omega\pm(\mu +\lambda))}\\
 {\supsub{B}{x}{N-1}}^{\pm}&=\lbr{\supsub{F}{x}{N-1}}^{\pm}\cos{(\omega\pm(\mu +\lambda))}\rbr
 \label{eq12}
 \end{align}
 \end{subequations}
 For clarity we shall now suppress the superscripts $(N-1)$. The system becomes 

  \begin{subequations}
  \begin{align}
0=&\dx\supN{F}{\pm}_{x}+ (m\pm \mu)\supN{F}{\pm}_{y}+ n\supN{F}{\pm}_{z} \label{eq13a}\\
\dx{\supN{G}{\pm}}=& -n \FO\supsub{F}{x}{\pm}-\frac{n}{2}\supsub{F}{x}{1}\supsub{F}{z}{N-2}\label{eq13b}\\
(m\pm\mu){\supN{G}{\pm}}=& +n \FO\supsub{F}{y}{\pm}+\frac{n}{2}\supsub{F}{y}{1}\supsub{F}{z}{N-2}\label{eq13c}\\
n {\supN{G}{\pm}}=& +n \FO\supsub{F}{z}{\pm}+ \FO_{,x}\supsub{F}{x}{\pm}\label{eq13d}\\
&+\frac{1}{2}\lbr \supN{F}{1}_x\dx\supsub{F}{z}{N-2}+\supsub{F}{z}{N-2}\lbr n\supN{F}{1}_z\pm\mu \supN{F}{1}_y\rbr\rbr\nonumber
  \end{align}
  \label{eq13}
  \end{subequations}
 From which it follows,
  \begin{align}
  \lbr(m\pm\mu)^2+n^2\rbr\supN{G}{\pm}=& -n\FO\dx\supN{F}{\pm}_{x}+n \FO_{,x}\supsub{F}{x}{\pm}\label{eq14}\\
  &+\frac{n}{2}\lbr\supsub{F}{z}{N-2}\lbr (m\pm 2\mu)\supN{F}{1}_y+n\supN{F}{1}_z\rbr+n \supN{F}{1}_x\dx\supsub{F}{z}{N-2}\rbr\nonumber
  \end{align}
Thus (\ref{eq13b}) and (\ref{eq14}) form a linear system in $\supsub{F}{z}{N-2},G^{(\pm)},F_x^{(\pm)}$. We next consider the system in order $N$. The system of equations will be the system (\ref{sys5}) with the index replaced by $(N)$ and additional sums of terms which are products of $\supN{\B}{j},\supN{\B}{N-j}, 0<j<N$ and their derivatives. After fourier analysis we find sums of terms of the form $T(x)\sin{(\mu y + \nu z +\lambda)}$. Clearly we may solve for the corresponding components of $\supN{\B}{N}$ provided $\nu \neq 0$. We must show that when $\nu=0$, we may choose $F_z^{(N-2)}$ appropriately so that the corresponding expression for $T(x)$ vanishes. We treat separately the $(\mu\neq 0,\nu =0)\: \& \:(\mu=\nu =0)$ cases. We start with $(\mu \neq 0, \nu=0)$. We replace $(m=\mu, n=0)$ in the system (\ref{sys5}) and we see that $F_y^{(N)}$ appears only in (\ref{5a}) and $F_x^{(N)}$ only in (\ref{5a},\ref{5d}). Thus, we focus only on (\ref{5b},\ref{5c}). If we can choose $F_z^{(N)}$ so that this system is solvable for $\Pi^{(N)}$ then we may determine $F_x^{(N)}$ from (\ref{5d}) and $F_y^{(N)}$ form  (\ref{5a}). Hence we focus on (\ref{5b},\ref{5c}).

The as yet unspecified function $F_z^{(N-2)}$ appears in order $N$ indirectly in $\B^{(N-1)}$, see (\ref{eq13}), and in the combination $\B^{(2)}\cdot \dl \B^{(N-2)}+\B^{(N-2)}\cdot \dl \B^{(2)}.$ The first term has no $x$ or $y$ component, while the second term either is non-resonant if it contains angular dependence on $y$ or $z$, or is zero if it depends on $x$. Thus, we consider only the $x$ and $y$ components of 
\begin{align}
\bm{V}=\B^{(1)}\cdot \dl \B^{(N-1)}+\B^{(N-1)}\cdot \dl \B^{(1)}.
\label{15}
\end{align}
Hence the system (\ref{5b},\ref{5c}) to $O(N)$ becomes 
\begin{subequations}
\begin{align}
\supN{\Pi}{N}_{,x}=\bm{V}\cdot \hat{x}= \B^{(1)}\cdot \dl {B}^{(N-1)}_x+\B^{(N-1)}\cdot \dl {B}^{(1)}_x \label{eq16a}\\
\supN{\Pi}{N}_{,y}=\bm{V}\cdot \hat{y}= \B^{(1)}\cdot \dl {B}^{(N-1)}_y+\B^{(N-1)}\cdot \dl {B}^{(1)}_y \label{eq16b}
\end{align}
\end{subequations}
a system for $G^{(N)}$ and $F_z^{(N-2)}$. This pair of equations is coupled to the system (\ref{eq13b},\ref{eq14}) for $(G^{\pm},F_x^{\pm})$.We focus on the structure of (\ref{eq16b}). We obtain easily that the z independent part of $\B^{(N-1)}\cdot \dl B_y^{(1)}$ and $\B^{(1)}\cdot \dl B_y^{(N-1)}$  are
\begin{subequations}
\begin{align}
\B^{(N-1)}\cdot \dl B_y^{(1)} = & \frac{1}{2}\sin{(\mu y +\lambda)}\left[\dx\supN{F}{1}_y\lbr\supN{F}{-}_x-\supN{F}{+}_x\rbr \right. \label{eq17a}\\
& \quad+ \left. m \supN{F}{1}_y\lbr\supN{F}{+}_y-\supN{F}{-}_y\rbr +n \supN{F}{1}_y\lbr\supN{F}{+}_z-\supN{F}{-}_z\rbr  \right]+\dots\nonumber\\
\B^{(1)}\cdot \dl B_y^{(N-1)} = & \frac{1}{2}\sin{(\mu y +\lambda)}\left[\supN{F}{1}_y\lbr-(m+\mu)\supN{F}{+}_y+(m-\mu)\supN{F}{-}_y\rbr \right.\label{eq17b}\\
& \quad+ \left.  \supN{F}{1}_x\dx\lbr\supN{F}{+}_y-\supN{F}{-}_y\rbr +n \supN{F}{1}_z\lbr\supN{F}{-}_y-\supN{F}{+}_y\rbr  \right]+\dots\nonumber
\end{align}
\end{subequations}
We follow one term of particular interest $\supN{F}{N-2}_{z,x}$. From (\ref{eq13c},\ref{eq13d}) 
\begin{subequations}
\begin{align}
\B^{(N-1)}\cdot \dl B_y^{(1)}+\B^{(1)}\cdot \dl B_y^{(N-1)} = \frac{m\mu}{n} \supN{F}{1}_y\frac{\supN{F}{1}_x}{\FO}\supN{F}{N-2}_{z,x}+\dots \label{18a}\end{align}
\end{subequations}
so that 
\begin{align}
\mu G^{(N)} =\supN{F}{1}_y\frac{\supN{F}{1}_x}{\FO}\supN{F}{N-2}_{z,x}+\dots\label{eq19}
\end{align}
Thus (\ref{eq19}) together with (\ref{eq13b},\ref{eq14}) for $(G^{(\pm)},F_x^{(\pm)})$ forms a closed system for the $\lbr \supsub{F}{z}{N-2},G^{(\pm)},F_x^{(\pm)} \rbr$ system. From (\ref{eq19}) , for ranges of values of $x$ for which neither $F_y^{(1)},F_x^{(1)}$ vanish, one may solve for the unknown functions.

 Next we must examine the case $\mu=0$ for which the general solution for the resonant mode is (see (\ref{eq8}))
 \begin{align}
 \B^{(N-2)}=\lbr 0,\supN{F}{N-2}_y(x), \supN{F}{N-2}_z(x)\rbr
 \label{eq20}
 \end{align}
 The angular dependence of the $\B^{(N-1)}$ system is similar to the $\B^{(1)}$ system since $\B^{(N-2)}$ is independent of the angles. Therefore, $$\B^{(N-1)}=\lbr \supsub{F}{x}{N-1}(x)\cos{\omega},\supsub{F}{y}{N-1}(x)\sin{\omega},\supsub{F}{z}{N-1}(x)\sin{\omega} \rbr ,\:\: \supsub{\Pi}{}{N-1}=\supN{G}{N-1}(x)\sin{\omega}$$ is 
 \begin{subequations}
 \begin{align}
 \dx\supN{F}{N-1}_{x}+&\: m \supN{F}{N-1}_y+n \supN{F}{N-1}_z=0\\
 \del_x \supN{G}{N-1} =&\: -n \supN{F}{0} \supN{F}{N-1}_{x}\label{21b}\\
m\supN{G}{N-1} =& \:n \supN{F}{0} \supsub{F}{y}{N-1}+ \supsub{F}{x}{1}\dx \supsub{F}{y}{N-2}\label{21c}\\
n\supN{G}{N-1} =& \:n \supN{F}{0} \supsub{F}{z}{N-1}+\del_x\supN{F}{0} \:\supsub{F}{x}{N-1} +\supsub{F}{x}{1}\dx \supsub{F}{z}{N-2}\\
&+ \supsub{F}{z}{1}\lbr n\supsub{F}{z}{N-2}+m\supsub{F}{y}{N-2}\rbr\nonumber
 \end{align}
 \label{eq21}
 \end{subequations}

If we now examine the $y$ component of $\B^{(N-1)}\cdot \dl \B^{(1)}$ and $\B^{(1)}\cdot \dl \B^{(N-1)}$ we see immediately that all the terms are proportional to $\sin{2\omega}$ and hence the $y$ and $z$ average is identically zero. Thus, this resonance if it occurs cannot be resolved. 
 However one should note that with stellarator symmetry for which every component of $\Pi,B_y,B_z$ is proportional to $\sin{(m y + n z+\lambda)}$ and $B_x$ is proportional to $\cos{(m y + n z+\lambda)}$ no constant terms appear and no $\mu=\nu=0$ resonance occurs. While this is scarcely a proof of the necessity of stellarator symmetry, it is a strong argument for its utility. As a counter-example to the need for stellarator symmetry, we note that the expression in \citep{weitzner2016expansions} would be possible without such symmetry.

 \section{Expansion in the distance from a flux surface for vacuum magnetic fields}
 \label{sec:Cauchy_vac}
We shall now investigate the compatibility conditions required for the existence of neighboring flux surfaces given consistent data on one flux surface. This problem is an analog of the Cauchy problem for vacuum magnetic fields with surfaces. We construct a local extension of the data in a region near the original surface on which data is given while maintaining global constraints of periodicity and the formal existence of flux surfaces.

In the following, we examine the possibility of expanding a vacuum magnetic field in a flat toroidal shell in a parameter which measures the distance from a magnetic flux surface. It was shown in \citep{weitzner2016expansions} that a broad class of such expansions are possible for an MHD equilibrium. Although not stated explicitly there a force-free field is also included in those results. The flux surface there was the plane $x=0$, and a particularly simple structure of the field on the surface was assumed. We explore the analogous case here. For reasons of consistency, we follow the notation of this paper instead of \citep{weitzner2016expansions}.
  A vacuum magnetic field which also lies on the magnetic surface $\psi(x,y,z)=$ constant, satisfies the constraint 
  \begin{align}
  \B=\dl\Phi = \dl \psi \times \dl \alpha
  \label{a1}
  \end{align}
 where $\Phi$ and $\alpha$ can be multivalued. If we introduce the angle coordinates on the ``torus'' $(\theta,\varphi)$ and 
 \begin{align}
 \cJ=\frac{\del(\psi,\theta,\varphi)}{\del(x,y,z)}
  \label{a2}
 \end{align}
 then (\ref{a1}) becomes $$\Phi_{,\psi}\dl \psi +\Phi_{,\theta}\dl \theta+\Phi_{,\varphi}\dl \varphi= \alpha_{,\theta} \dl \psi\times \dl \theta +\alpha_{,\varphi} \dl \psi\times \dl \varphi. $$ Dotting with $\dl\psi, (\dl\varphi\times\dl\psi) $ and $(\dl\psi\times\dl\theta)$ we get
 \begin{subequations}
 \begin{align}
 \Phi_{,\psi} |\dl\psi|^2+\Phi_{,\theta}\dl\psi\cdot\dl\theta+\Phi_{,\varphi}\dl\psi\cdot\dl\varphi=0 \label{a3}\\
 \cJ\Phi_{,\theta}= -\dl \psi\times \dl\alpha \cdot \dl \psi\times \dl\varphi
\label{a4}\\
\cJ\Phi_{,\varphi}= \dl \psi\times \dl\alpha \cdot \dl \psi\times \dl\theta 
\label{a5}
 \end{align}
 \label{sysa3a4a5}
 \end{subequations}
 the latter two equations are exactly those found in \citep{weitzner2014ideal,sengupta_weitzner2018}. We may rewrite (\ref{a3}) as 
 \begin{align}
 \cJ \Phi_{,\psi}+\dl \psi\times \dl\alpha \cdot \dl \varphi\times \dl \theta =0
 \label{a6}
 \end{align}
 Thus, a vacuum field with surfaces is characterized by the system (\ref{a4},\ref{a5},\ref{a6}). We now specialize to the flat toroidal shell in three dimensions in which we take $y$ and $z$ as the two angles and assume that physical quantities are $2\pi$ periodic. The system reduces to
  \begin{subequations}
 \begin{align}
\psi_{,x}\begin{pmatrix}
\Phi_{,y}\\ \Phi_{,z}
\end{pmatrix}&=
\begin{pmatrix}
(\psi_{,x}^2+\psi_{,y}^2) \quad  \quad \psi_{,y}\psi_{,z}\\ 
\quad \psi_{,y}\psi_{,z} \quad  \quad (\psi_{,x}^2+\psi_{,z}^2)
\end{pmatrix}
\begin{pmatrix}
-\alpha_{,z}\\ \alpha_{,y}
\end{pmatrix}
\label{a8}\\
\psi_{,x}\Phi_{,\psi}&=-(\psi_{,z}\alpha_{,y}-\psi_{,y}\alpha_{,z}) \label{a9}.
\end{align}
\label{sysa7a8a9}
\end{subequations}
 Finally, it is convenient to use the inverse representation and assume $x=x(\psi,y,z)$ rather than $\psi=\psi(x,y,z)$
\begin{subequations}
 \begin{align}
x_{,\psi}\begin{pmatrix}
\Phi_{,y}\\ \Phi_{,z}
\end{pmatrix}&=
\begin{pmatrix}
(1+x_{,y}^2) \quad  \quad x_{,y}x_{,z}\\ 
\quad x_{,y}x_{,z} \quad  \quad (1+x_{,z}^2)
\end{pmatrix}
\begin{pmatrix}
-\alpha_{,z}\\ \alpha_{,y}
\end{pmatrix}
\label{a10a11}\\
\Phi_{,\psi}&=(x_{,z}\alpha_{,y}-x_{,y}\alpha_{,z}) =\xcpsi\BD x\label{a12}\:\:
\end{align}
\label{sysa10a11a12}
\end{subequations}
where the magnetic differential operator $(\BD)$ in the inverse representation is given by
\begin{align}
\BD=\frac{1}{x_{,\psi}}\lbr \alpha_{,y}\dz- \alpha_{,z}\dy \rbr\:.
\label{BDinvI}
\end{align}
The first two equations are exactly those for a MHD equilibrium (see Eqn (25) from \cite{sengupta_weitzner2018}). They constitute the pseudo-analytic Cauchy-Riemann equation for the $(\Phi,\alpha)$ system on a constant $\psi$ flux surface, assuming also that $\xcpsi$ is known. Since $\xcpsi$ is not an intrinsic surface function alone, properties of nearby surfaces appear in the system.  The third equation is similar to pressure balance c.f Eqn (28) from \citep{weitzner2016expansions} and Eqn (27) from \citep{sengupta_weitzner2018}. However, the difference is substantial. For a force-free field one would add an arbitrary flux function $F(\psi)$ to (\ref{a12}). It will become clear that this omission appropriate to a vacuum magnetic field produces fundamental differences.

Using (\ref{a12}) we can rewrite (\ref{a10a11},\ref{a12}) as
\begin{align}
\begin{cases}
x_{,\psi}\Phi_{,y}-x_{,y}\Phi_{,\psi}= -\alpha_{,z} \nonumber\\
x_{,\psi}\Phi_{,z}-x_{,z}\Phi_{,\psi}= +\alpha_{,y}
\label{a10a11*} \tag{\ref{a10a11}*}
\end{cases}
\\
\Phicpsi=\frac{\xcpsi}{1+\xcy^2 + \xcz^2}\lbr \xcz \Phicz +\xcy \Phicy\rbr
\label{phipsi2} \tag{\ref{a12}*}
\end{align}
We note that Eqn. (\ref{phipsi2}) is exactly in the form of a Cauchy problem: given the derivatives of $\Phi$ on a constant $\psi$ surface along with the coefficients that involve all the three derivatives of $x$, can we find $\Phi$ in the neighbourhood of the surface? A straight forward application of Cauchy-Kowalevski theorem is not possible because the data is given on a constant $\psi$ which is also a characteristic surface of the system. It is well known that if the data is given on a characteristic surface, certain additional solvability conditions need to be satisfied. In the following we shall discuss these conditions and construct an expansion of vacuum fields in the vicinity of the surface $x=0$, which is also assumed to be a flux surface.

\subsection{Series expansion in $\psi$ about a given flux surface}
 We may use the system (\ref{a10a11},\ref{a12}) or equivalently (\ref{a10a11*},\ref{a12}) to explore the possibility of a formal power series expansion of $(x(\psi,y,z),\Phi(x,y,z),\alpha(\psi,y,z))$ in the coordinate $\psi$. The expansion repeats the process laid out in \citep{weitzner2016expansions}. We expand as 
 \begin{align}
 (x,\Phi,\alpha)=\sum_{n>0}(x^{(n)}(y,z),\Phi^{(n)}(y,z),\alpha^{(n)}(y,z)) \psi^n 
 \label{a13}
 \end{align}
 We assume that $x=0$ is a magnetic surface so that $x^{(0)}=0$. We find easily to lowest order that
 \begin{subequations}
 \begin{align}
 x^{(1)}\Phi^{(0)}_{,y} &= -\alpha^{(0)}_{,z} \label{a14a}\\
 x^{(1)}\Phi^{(0)}_{,z} &= +\alpha^{(0)}_{,y}\label{a14b}\\
 \Phi^{(1)}=&0 \label{a14c}
 \end{align}
 \label{ZeroO}
 \end{subequations}
From the lowest order generalized CR conditions (\ref{a14a},\ref{a14b}), the following orthogonality condition is obtained
\begin{align}
\PhiOy\alphaOy+\PhiOz\alphaOz=0.
\label{ortho0}
\end{align}
To first order, we have
\begin{subequations}
 \begin{align}
 2 x^{(2)}\Phi^{(0)}_{,y} &= -\alpha^{(1)}_{,z} \label{a15a}\\
 2 x^{(2)}\Phi^{(0)}_{,z} &= +\alpha^{(1)}_{,y}\label{a15b}\\
 2 \Phi^{(2)}=& \lbr \alpha^{(0)}_{,y}\dz -\alpha^{(0)}_{,z}\dy \rbr\xOne\:. \label{a15c}
 \end{align}
  \label{1stO}
 \end{subequations}
 Eliminating $\xTwo$ between (\ref{a15a},\ref{a15b}) we obtain a homogeneous magnetic differential equation for $\alphaOne$
 \begin{align}
  \alpha^{(0)}_{,y}\alpha^{(1)}_{,z}-\alpha^{(0)}_{,z}\alpha^{(1)}_{,y}=0 ,
 \label{MDEalpha1}
 \end{align}
which implies that $ \alphaOne$ is a function of $\alphaO$ i.e. $\alphaOne=\alphabarOne(\alphaO)$. Therefore, $\xTwo$ has the form
\begin{align}
\xTwo=\frac{1}{2}\xOne\: \alphabarPOne (\alphaO).
\label{x2}
\end{align}
We only require $\alphabarPOne (\alphaO)$ to be a nonzero even function of $\alphaO$. Going to the second order and using the lower order equations, we find
\begin{subequations}
 \begin{align}
-\alpha^{(2)}_{,z} &=  3 x^{(3)}\Phi^{(0)}_{,y} + x^{(1)}\Phi^{(2)}_{,y}-2x^{(1)}\Phi^{(2)}_{,y} \label{a16a}\\
 +\alpha^{(2)}_{,y} &= 3 x^{(3)}\Phi^{(0)}_{,z} +x^{(1)}\Phi^{(2)}_{,z}-2x^{(1)}\Phi^{(2)}_{,z}\label{a16b}\\
  \Phi^{(3)}&= \lbr \alpha^{(0)}_{,y}\dz -\alpha^{(0)}_{,z}\dy \rbr \lbr \xOne\:\alphabarPOne \rbr \:.\label{a16c}
 \end{align}
 \label{2ndO}
 \end{subequations}
As before, eliminating $x^{(3)}$ from (\ref{a16a},\ref{a16b}) we obtain a magnetic differential equation for $\alphaTwo$
\begin{align}
\BOD\alphaTwo =\lbr \xOne \rbr^2 \lbr \alpha^{(0)}_{,z}\dz +\alpha^{(0)}_{,y}\dy \rbr\lbr \frac{\PhiTwo}{\lbr \xOne \rbr^2 }\rbr
\label{MDEalpha2}
\end{align}
where,
\begin{align}
\BOD=\frac{1}{\xOne}\lbr \alpha^{(0)}_{,y}\dz -\alpha^{(0)}_{,z}\dy \rbr .
\label{BODinvI}
\end{align}
From (\ref{a14a},\ref{a14b}) and  (\ref{a15c}) it follows that
 \begin{subequations}
 \begin{align}
 \frac{2\PhiTwo}{(\xOne)^2} &= -\lbr \del_{yy}+\del_{zz} \rbr\PhiO \label{I2}\\
\lbr\alpha^{(0)}_{,y}\dz -\alpha^{(0)}_{,z}\dy \rbr \alpha^{(2)} &=- \frac{(\xOne)^3}{2}\lbr \alpha^{(0)}_y \del_y+\alpha^{(0)}_z \del_z \rbr \lbr \del_{yy}+\del_{zz} \rbr\PhiO \label{MDE2alpha2}
 \end{align}
 \end{subequations}
 We focus on the relation (\ref{MDE2alpha2}). Since $\alpha^{(n)}$ for all $n$ must have the form $$\alpha^{(n)} = f^{(n)}y +g^{(n)} z + \tilde{\alpha}^{(n)}(y,z)$$ where $(f^{(n)},g^{(n)})$ are constants and $\tilde{\alpha}^{(n)}$ is a periodic in $y$ and $z$, it follows that the integral on the left side of (\ref{MDE2alpha2}) along the lowest order closed magnetic line must be a constant independent of the magnetic line in question. The detailed arguments are given in \citep{weitzner2016expansions}, where it was shown that for an ideal MHD equilibrium or a force-free field with a particular structure of $\alpha^{(0)}(y,z)$ (denoted by $\psi^{(0)}$ in \cite{weitzner2016expansions}) this condition is satisfied and that the expansion can be carried to all orders in $\psi$ (denoted by $\gamma$ in \cite{weitzner2016expansions}).
 
The solvability condition for an equation of the form 
\begin{align}
\lbr\alpha^{(0)}_{,y}\dz -\alpha^{(0)}_{,z}\dy \rbr \alpha^{(2)} =I(y,z) \nonumber\\
\text{is}\quad \oint_{\alphaO} \frac{dy }{\alpha^{(0)}_{,z}}I(y,z)=-\oint_{\alphaO}  \frac{dz}{\alpha^{(0)}_{,y}}I(y,z) = \text{constant},
\label{MDEsolvability}
\end{align}
 where the the integrals are evaluated along a fixed field line label $\alphaO(y,z)=\alphaO$ and the constant must be independent of $\alphaO$. We can use implicit function theorem to express one of the angles in terms of the other angle and $\alphaO$. Upon carrying out the integral in (\ref{MDEsolvability}) we are left with a periodic function of $\alphaO$, since all the quantities that appear in the integrand are periodic functions. Only under very restrictive conditions would this periodic function of $\alphaO$ be a constant as is required by the solvability constraint. In ideal MHD equilibrium of force-free field there are free functions of the form $\overline{\Phi}(\alphaO)$ (cf. Eqn. (26) in \citep{weitzner2016expansions}) which can be chosen so that the overall integral is a constant, but in the vacuum field case, no such free functions are available and therefore the solvability condition is not easily satisfied. 
 
  We now assume the same lowest order structure for $\alpha^{(0)}$ as used previously and we examine the differences between vacuum fields and equilibrium. Specifically we choose 
 \begin{subequations}
 \begin{align}
 \alpha^{(0)}&= \mu(y)-\nu(z)
 \label{19a}\\
 \quad  \mu(y)&=m y +P(y),\quad \nu(z)=n z +Q(z)
 \label{19bc}
 \end{align}
 \end{subequations}
 where P and Q are arbitrary $2\pi$ periodic functions in their arguments and the integers $(m,n)$ are relatively prime and equilibria were easily found in \citep{weitzner2016expansions}. It follows 
 \begin{subequations}
 \begin{align}
 x^{(1)}(y,z)&=\mu'(y)\nu'(z)\\
 \Phi^{(0)}(y,z)&=\int\frac{dy}{\mu'(y)}+\int\frac{dz}{\nu'(z)}
 \end{align}
 \label{sysa20ab}
 \end{subequations}
The integral constraint (\ref{MDEsolvability}) then becomes
\begin{align}
\oint_{\alphaO} dz \lbr \mu' \nu' \rbr^3 \del_{yy}\lbr\frac{1}{\mu'}\rbr - \oint_{\alphaO} dy \lbr \mu' \nu' \rbr^3 \del_{zz}\lbr\frac{1}{\nu'}\rbr = \text{constant}
\label{I2munu}
\end{align}
As discussed earlier, for general $(m,n,P,Q)$, the left side must be a periodic function of the field line label $\alphaO$ which in general will not be a constant. However, in the following cases we can indeed satisfy the constraint:

\begin{enumerate}
\item \quad Symmetric solution when ($Q=0$ or $P=0$)\\
\item \quad Non-symmetric solution when ($m=n$) and $(P,Q)$ have the same functional form. 
\end{enumerate}
In the symmetric solution if either $\mu'$ or $\nu'$ is a constant then one of the integrals drop out and the other one is a constant. In the more interesting non-symmetric case, the integrals cancel each other because of the symmetry of the $(\mu',\nu')$ terms in the integrals. It is to be noted that in this case although $(m=n)$, we do not have a helically symmetric solution because $\alphaO=m(y-z)+P(y)-P(z)$ is not expressible through a single helical angle for an arbitrary periodic function $P(y)$. Such solutions also appear in ideal MHD and has been in experiments \citep{gill1992snake}. They are highly localized near a rational surface (in most cases near $\iota=1$) and they vary very little along the closed magnetic field line. An analytical reduced MHD description of a tokamak MHD ``snake'' is provided in \cite{senguptaHasssam2017subalf}. Numerical solutions and typical pictures of a tokamak ``snake-like'' MHD equilibrium can be found in \citep{cooper2011jetsnake,cooper2011helical,sugiyama2013snakes}. In the next section we report such ``snake-like'' states in vacuum magnetic fields and show that the expansion can be carried out to all orders. 
 
\subsection{Vacuum ``snake-like" states}
We have seen previously that with the choice $\mu(y)=\nu(y),$ we can satisfy the solvability constraint. Now we proceed to show that one can construct such an expansion to all orders. The methodology is similar to \citep{weitzner2016expansions} but there are significant differences. We return to Eqn. (\ref{MDE2alpha2}) and see that the solvability condition is satisfied because both the integrands in (\ref{I2munu}) are even under the exchange of the angles $(y,z)$. Therefore, an exchange of the integration variables lead to a cancellation of the two integrals. We can then proceed to solve for $\alphaTwo$ in the form
\begin{align}
\alphaTwo = \alphabarTwo(\alphaO)+\aav{\alphaTwo}{},
\label{alphaTwosol}
\end{align}
where $\alphabarTwo$ and $\aav{\alphaTwo}{}$ are respectively the homogeneous and the inhomogeneous solutions of Eqn. (\ref{MDE2alpha2}). We can now solve for $\xThree$ from Eqn.(\ref{a16a}) 
\begin{align}
3\lbr \frac{\xThree}{\xOne}\rbr=-\alphabarPTwo-\frac{\del_z\aav{\alphaTwo}{}}{\alphaOz}+\frac{\lbr\xOne\rbr^3}{\alphaOz}\del_y\lbr \del_{yy}+\del_{zz}\rbr\PhiO
\label{xThreesol}
\end{align}
We now proceed to the third order and after some manipulation, obtain the following 
\begin{subequations}
\begin{align}
4\PhiFour&=3\lbr \frac{\xThree}{\xOne}\rbr \lbr \alphaOy \dz -\alphaOz \dy\rbr\xOne-\frac{1}{2}(\xOne)^3(\xOne_{,y}\dy +\xOne_{,z}\dz)\lbr  \del_{yy}+\del_{zz}\rbr\PhiO
\label{Phi4}\\
-\alphaThree_{,z}&=4 \xFour \PhiOy + 2\lbr\xTwo \rbr^2 \del_y\lbr \frac{\PhiTwo}{\xTwo}\rbr+\lbr\xOne \rbr^2 \del_y\lbr \frac{\PhiThree}{\lbr\xOne\rbr^3}\rbr \label{alf3z}\\
+\alphaThree_{,y}&=4 \xFour \PhiOz + 2\lbr\xTwo \rbr^2 \del_z\lbr \frac{\PhiTwo}{\xTwo}\rbr+\lbr\xOne \rbr^4 \del_z\lbr \frac{\PhiThree}{\lbr\xOne\rbr^3}\rbr \label{alf3y} \:.
\end{align}
\end{subequations}
From Eqns. (\ref{ZeroO}) and (\ref{1stO},\ref{2ndO}) it follows
\begin{align}
\lbr\frac{\PhiTwo}{\xTwo}\rbr=-\lbr\frac{\xOne}{\alphabarPOne}\rbr\lbr  \del_{yy}+\del_{zz}\rbr\PhiO, \quad \lbr \frac{\PhiThree}{\lbr\xOne\rbr^3}\rbr=-\lbr\frac{\xOne}{\alphabarPOne}\rbr^{-1}\lbr  \del_{yy}+\del_{zz}\rbr\PhiO.\label{phi2phi3sol}
\end{align}
The magnetic differential equation for $\alphaThree$ is then obtained after eliminating $\xFour$ from Eqns.(\ref{alf3z},\ref{alf3y}) 
\begin{align}
\lbr \alphaOy \dz -\alphaOz \dy\rbr \alphaThree =&-2\lbr\xTwo \rbr^2\lbr\alpha^{(0)}_{,z}\dz +\alpha^{(0)}_{,y}\dy \rbr \lbr\frac{\PhiTwo}{\xTwo}\rbr \nonumber\\
&-\lbr\xOne \rbr^4 \lbr\alpha^{(0)}_{,z}\dz +\alpha^{(0)}_{,y}\dy \rbr \lbr \frac{\PhiThree}{\lbr\xOne\rbr^3}\rbr
\label{MDEalphaThree}
\end{align}
We now note that for the ``snake'' solution we have the following
\begin{align}
&\xOne=\mu'(y)\mu'(z)\:,\quad  \alphaO =\mu(y)-\mu(z)\:,\quad \PhiO=\int \frac{dz}{\mu'(z)}+\int \frac{dy}{\mu'(y)}
\\
&\lbr \alphaOy \dz -\alphaOz \dy\rbr =\mu'(y)\dz +\mu'(z)\dy\:\:, \quad \lbr\alpha^{(0)}_{,z}\dz +\alpha^{(0)}_{,y}\dy \rbr =\mu'(y)\dy -\mu'(z)\dz\nonumber
\end{align}
Since $\alphabarPOne (\alphaO)$ is assumed to be an even function of $\alphaO$, we observe from (\ref{phi2phi3sol}) that the two functions involving $(\PhiTwo,\PhiThree)$ are even under the exchange of $y$ and $z$ and so is $(\xOne,\xTwo,\PhiO)$ , while the operator $\lbr\alpha^{(0)}_{,z}\dz +\alpha^{(0)}_{,y}\dy \rbr$ is odd. Therefore, the solvability condition (\ref{MDEsolvability}) for $\alphaThree$ is identically satisfied because of the odd nature of the expression on the right side of (\ref{MDEalphaThree}). Finally, we can solve for $\alphaThree$ in the form similar to that of $\alphaTwo$ in Eqn. (\ref{alphaTwosol}), i.e
\begin{align}
\alphaThree = \alphabarThree(\alphaO)+\aav{\alphaThree}{},
\label{alphaThreesol}
\end{align}
Thus, we obtain three ``free-functions'' ($\alphabarOne,\alphabarTwo,\alphabarThree$), all functions of $\alphaO$ that can be used to satisfy the consistency conditions occurring in higher orders. We shall now work towards an inductive proof similar to \citep{weitzner2016expansions}. In the following analysis we shall be using the form of equations given by (\ref{a10a11*},\ref{a12}). 

Let us begin with the even order $O\lbr\psi^{(2n-2)}\rbr$ with $n\geq 1$. We shall only keep relevant terms and subsume the rest in dots. The generalized CR equations to this order are
\begin{align}
-\supsub{\alpha}{,z}{2n-2}&=(2n-1)\supsub{x}{}{2n-1}\PhiOy+.... \nonumber\\
\supsub{\alpha}{,y}{2n-2}&=(2n-1)\supsub{x}{}{2n-1}\PhiOz+....
\label{Opsi2nm2}
\end{align}
We shall assume that the magnetic differential equation for $\supsub{\alpha}{}{2n-2}$ can be satisfied through the use of a lower order ``free-function''. We then express the solution for $\supsub{\alpha}{}{2n-2}$ in the form
\begin{align}
\supsub{\alpha}{}{2n-2} = \supbsub{\alpha}{}{2n-2}(\alphaO)+\aav{\supsub{\alpha}{}{2n-2}}{}.
\label{alpha2nm2sol}
\end{align}
Solution for $\supsub{x}{}{2n-1}$ using (\ref{Opsi2nm2},\ref{alpha2nm2sol}) takes the form
\begin{align}
(2n-1)\frac{\supsub{x}{}{2n-1}}{\xOne}= \supbPsub{\alpha}{}{2n-2}(\alphaO) +\frac{\del_z\aav{\supsub{\alpha}{}{2n-2}}{}}{\alphaOz}+....
\label{x2nm1sol}
\end{align}
Proceeding to $O(\psi^{(2n-1)})$ and using (\ref{Opsi2nm2},\ref{x2nm1sol}) we find
\begin{align}
2n\:\Phi^{(2n)}&= \lbr\supsub{\alpha}{,y}{2n-2} \del_z -\supsub{\alpha}{,z}{2n-2}\del_y \rbr\xOne +\lbr \alphaOy\del_z -\alphaOz \del_y \rbr \supsub{x}{}{2n-1}+..\nonumber\\
&=\frac{2n}{2n-1}\supbPsub{\alpha}{}{2n-2}\lbr \alphaOy\del_z -\alphaOz \del_y \rbr \xOne +...
\label{Phi2n}
\end{align}
The CR equation pair to this order is given by
\begin{align}
-\supsub{\alpha}{,z}{2n-1}&=(2n)\supsub{x}{}{2n}\PhiOy+.... \nonumber\\
\supsub{\alpha}{,y}{2n-1}&=(2n)\supsub{x}{}{2n}\PhiOz+....
\label{Opsi2nm1}
\end{align}
We shall once again assume that the magnetic differential equation for $\supsub{\alpha}{}{2n-1}$, obtained by eliminating $\supsub{x}{}{2n}$ can be solved. This yields
\begin{align}
\supsub{\alpha}{}{2n-1} &= \supbsub{\alpha}{}{2n-1}(\alphaO)+\aav{\supsub{\alpha}{}{2n-1}}{} \nonumber\\
(2n)\frac{\supsub{x}{}{2n}}{\xOne}&= \supbPsub{\alpha}{}{2n-1}(\alphaO) +\frac{\del_z\aav{\supsub{\alpha}{}{2n-1}}{}}{\alphaOz}+....
\label{Oenm1sols}
\end{align}
Next we proceed to $O(\psi^{(2n)})$. The CR equations are 
\begin{align}
-\supsub{\alpha}{,z}{2n}&=(2n+1)\supsub{x}{}{2n+1}\PhiOy+\lbr\xOne\rbr^{(2n+1)}\del_y\lbr \frac{\Phi^{(2n)}}{\lbr \xOne\rbr^{(2n)}}\rbr+.... \nonumber\\
\supsub{\alpha}{,y}{2n}&=(2n+1)\supsub{x}{}{2n+1}\PhiOz+\lbr\xOne\rbr^{(2n+1)}\del_z\lbr \frac{\Phi^{(2n)}}{\lbr \xOne\rbr^{(2n)}}\rbr+....
\label{Opsi2n}
\end{align}
The magnetic differential equation for $\supsub{\alpha}{}{2n}$ is
\begin{align}
\lbr \alphaOy \dz -\alphaOz \dy\rbr \supsub{\alpha}{}{2n}=& -\lbr\xOne \rbr^{(2n+1)}\lbr\alphaOz\dz +\alphaOy\dy \rbr \lbr\frac{\phi^{(2n)}}{\lbr \xOne \rbr^{(2n)}}\rbr +...
\label{MDEalpha2n}
\end{align}
From (\ref{Phi2n}) it follows that we can use the ``free-function'', $\supbPsub{\alpha}{}{2n-2}$ to satisfy the consistency condition for Eqn. (\ref{MDEalpha2n}). Therefore, we can solve for  $(\supsub{\alpha}{}{2n},\supsub{x}{}{2n+1})$ in a form analogous to (\ref{Oenm1sols}) with a new ``free-function'' $\supbsub{\alpha}{}{2n}$. We also obtain from this order $\Phi^{(2n+1)}$ in the form
\begin{align}
(2n+1)\Phi^{(2n+1)}&= \frac{2n+1}{2n}\supbPsub{\alpha}{}{2n-1}\lbr \alphaOy\del_z -\alphaOz \del_y \rbr \xOne +...
\label{Phi2np1}
\end{align}
It is now clear that when we proceed to $O(\psi^{(2n+1)})$, the corresponding magnetic differential equation for 
$\supsub{\alpha}{}{2n+1}$ will be exactly of the form (\ref{MDEalpha2n}) with ($2n$) replaced by $(2n+1)$. We can then use  $\supbPsub{\alpha}{}{2n-1}$ from $\Phi^{(2n+1)}$ to satisfy the consistency condition. Once the solvability condition is satisfied we can solve for $\supsub{\alpha}{}{2n+1}$ exactly like the previous orders in terms of a new ``free-function'' $\supbsub{\alpha}{}{2n+1}$.

Therefore, if we assume that the system is solvable for $(\supsub{\alpha}{}{2n-2}$,$\supsub{\alpha}{}{2n-1})$ due to lower order ``free-functions'', we obtain two functions ($\supbsub{\alpha}{}{2n-2},\supbsub{\alpha}{}{2n-1}$) which can be used to satisfy the consistency conditions to order $(2n)$ and $(2n+1)$. At the end of our analysis, we find two new functions ($\supbsub{\alpha}{}{2n},\supbsub{\alpha}{}{2n+1}$) that can be used to satisfy higher order consistency conditions. Since in our discussion $n$ is arbitrary and we have already demonstrated consistency for $(n=1)$, we can use induction to prove that the vacuum ``snake'' equilibrium expansion can be carried out to all orders. The question of convergence of the formal series expansion is beyond the scope of the present work.

\section{Discussion}
\label{sec:conclusions}

In this work, we have addressed the question of the existence of low-shear non-symmetric vacuum magnetic fields and MHD equilibria with nested flux surfaces in a flat toroidal shell. We have assumed a low-shear system with closed field lines for which KAM theory is not directly applicable. Closed field line toroidal systems are known to be highly constrained compared to systems with a continuous spatial symmetry. Periodicity requirement imposes specific structures in such a system that has been previously shown to be conducive to carrying out formal perturbation expansions to all orders.  We have considered two different perturbation expansions in this paper. The first approach is based on the amplitude of non-symmetric vacuum and equilibrium magnetic fields added to a simple closed field-line system on a torus. The second one is an expansion in the distance from a flux surface. Similar expansions for MHD (and force-free) equilibrium were carried out in \citep{weitzner2014ideal,weitzner2016expansions} and the present work extends these ideas to the case of vacuum magnetic field. We have not attempted to prove convergence of the series but have presented the general structure of the perturbation series of both of these expansions. We shall now discuss the two approaches and point out the main differences we have found between vacuum fields and equilibrium (or force-free) fields.

In the amplitude expansion analysis presented in section \ref{sec:straightB0}, we have shown that starting with a simple closed magnetic field $\BO=\zh$, in a flat toroidal shell with $(y,z)$ as the two angles, we can ensure persistence of nested flux surfaces order by order, provided two conditions be satisfied at each order. The first condition is that all non-symmetric vacuum field perturbations at order $n$ be of the form $\dz H^{(n)}$, where $H^{(n)}(x,y,z)$ is a harmonic function. The second necessary condition is that at each order a certain integral involving lower order quantities must be a constant. Systems with stellarator symmetry are shown to satisfy these conditions to all orders. 

We then revisited the amplitude expansion in an ideal MHD system but for very low magnetic shear. Using a slightly different formalism compared to \citep{weitzner2014ideal}, we show how addition of specific resonant harmonics to lower orders can eliminate resonances to higher order. We have allowed a non-constant lowest order magnetic field $\BO(x)$ and finite pressure $\pO(x)$. We have shown that if a perturbation of the form $\supN{\B}{1}=(\supsub{F}{x}{1}(x)\cos{\omega},\supsub{F}{y}{1}(x)\sin{\omega},\supsub{F}{z}{1}(x)\sin{\omega})\:,\: \supN{\Pi}{1}=\supN{G}{1}(x)\sin{\omega}
$, where $\omega= m y + n z$ is added to first order, resonant quantities take two distinct forms: $\B^{(N)}=(0,0,\supsub{F}{z}{N})\cos{(m y +\lambda)}$ for $(m\neq 0,n=0)$ and $\B^{(N)}=(0,\supsub{F}{y}{N}(x),\supsub{F}{z}{N}(x))$ for  $(m=n=0)$. We have shown that is possible to eliminate resonance at $O(N)$ of the first form through addition of $ \supN{\B}{N-2}=(0,0,\supsub{F}{z}{N-2})\cos{(\mu y +\lambda)},
$ at order $O(N-2)$. A similar strategy is shown to not work for the resonance of the second kind. However, for  stellarator symmetric systems, the resonance of the second type does not appear at all. This once again demonstrates the importance of stellarator symmetry.

We have then addressed the construction of neighboring flux surfaces for a vacuum magnetic field with data given on one flux surface. We have noted a fundamental difference in the structure of force-free and ideal MHD equilibrium and vacuum magnetic fields. The latter is a third order system whereas the other two are fourth order systems. The absence of one of the real characteristics in a vacuum field implies that constraints like constancy of the $\oint dl/B$ integral on a flux surface never appear for vacuum fields. This lack of constraint might give the impression that constructing vacuum magnetic fields with surfaces is easier than constructing a low beta MHD equilibrium with closed field lines. However, through an explicit construction of a perturbation series in the distance from a flux surface, we have shown in section (\ref{sec:Cauchy_vac}) that it is harder to find vacuum fields with surfaces than force-free or MHD equilibrium. We have identified a nontrivial constraint that must be satisfied by the lowest order quantities defined on the flux surface. The same formalism that provided a broad class of self-consistent ideal MHD solutions in \citep{weitzner2016expansions} failed to yield self-consistent vacuum fields in general. Only one highly constrained perturbative solution analogous to a tokamak ``snake-like'' equilibrium \citep{cooper2011jetsnake,senguptaHasssam2017subalf}, was shown to exist. 
 
 The formal proof of carrying out the series to all orders is also more involved than that in ideal MHD. The main difference arises because of the unavailability of a free-function which is a function of the lowest order field-line label for vacuum fields, which leads to a change in the structure of the perturbation expansion. We need to consider two successive orders at a time in our analysis unlike \citep{weitzner2016expansions}. 

More work is needed to generalize our analysis of vacuum and equilibrium magnetic fields on a flat toroidal shell to a real torus. In particular, it is of importance to understand the general form of the constraints that we encountered in this work. For the amplitude expansion case, we saw that a discrete symmetry like stellarator symmetry could satisfy the constraints. We remind the reader that while stellarator symmetry was used to verify construction of solutions to all orders in many of the expansions, it was not necessary for the expansion of an equilibrium given data on an initial flux surface \citep{weitzner2016expansions}. It would be of interest to check what (if any) other classes of solutions that break stellarator symmetry, can still satisfy the constraints. Similarly, we would like to investigate if more general non-symmetric vacuum magnetic field expansions besides the ``snake-like'' equilibrium exist. 

W. S. acknowledges stimulating discussion with J. W. Burby, A. B. Hassam, and M. Landreman. We also thank the reviewers for their constructive suggestions. This research was funded by the US DOE Grant No. DEFG02-86ER53223.

\appendix

\section{KAM theory for low-shear (properly degenerate systems)} 
\label{App:AppendixA}
Using the notations of this paper, we shall now rephrase two theorems due to Arnold \citep[theorem 16 and 17 and the remark afterward, p.~187]{arnold_dynamical_iii}. Arnold showed that for low-shear Hamiltonians of $1\sfrac{1}{2}$ and two degrees of freedom are ``more integrable" than the usual perturbed system in the following sense:
\begin{theorem}  \label{thm1}
    In a properly degenerate (low-shear) system with $1\dfrac{1}{2}$ and two degrees of freedom for all initial conditions, the values of ``action" (flux-surface $\psi$) variable remain forever near their initial values.
\end{theorem}

\begin{theorem} \label{thm2}
    Suppose a degenerate (low shear) system is made non-degenerate through perturbation, then the measure of the set of tori disappearing under the perturbation is exponentially small $O(\exp{(-\text{const}/\ep)})$ and the deviation of the torus from an unperturbed surface is $O(\ep)$.
\end{theorem}

 The degeneracy condition (low-shear) can be further relaxed \citep{pyartli1969diophantine,sevryuk1995kam,chierchia1982smooth} and so long as the magnetic shear is small but non-zero e.g. $( \alphabarPOne(\alphaO)\neq 0)$, invariant tori continue to exist. The exponentially small splitting of separatrices have been also shown in \citep{holmes1988exponentially, delshams1998exponentially}.

\bibliographystyle{jpp.bst}

\bibliography{plasmalit}

\end{document}